\documentclass[acmsmall]{acmart}
\AtBeginEnvironment{quote}{\smaller}
\usepackage{footnote}
\usepackage{balance}
\usepackage{url}
\usepackage{caption}
\usepackage{subcaption}
\usepackage{enumitem}

\AtBeginDocument{%
  \providecommand\BibTeX{{%
    \normalfont B\kern-0.5em{\scshape i\kern-0.25em b}\kern-0.8em\TeX}}}

\begin{document}

\newcommand{\highlight}[1]{\textcolor{blue}{#1}}  

\title[Politics of Fear]{The Politics of Fear and the Experience of Bangladeshi Religious Minority Communities Using Social Media Platforms}

\setcopyright{acmlicensed}
\acmJournal{PACMHCI}
\acmYear{2024} \acmVolume{8} \acmNumber{CSCW2} \acmArticle{387}
\acmMonth{11}\acmDOI{10.1145/3686926}


\author{Mohammad Rashidujjaman Rifat} \authornote{Both authors contributed equally to this paper}
 \affiliation{
   \department{Department of Computer Science}
   \institution{University of Toronto}
   \city{Toronto}
   \state{Ontario}
   \country{Canada}
 }
 \email{rifat@cs.toronto.edu}

 \author{Dipto Das} \authornotemark[1]
 \affiliation{%
   \department{Department of Computer Science}
   \institution{University of Toronto}
   \city{Toronto}
   \state{Ontario}
   \country{Canada}
 }
 \email{diptodas@cs.toronto.edu}

 \author{Arpon Podder}
 \affiliation{%
   \institution{Khulna University of Engineering \& Technology}
   \city{Khulna}
   \country{Bangladesh}
 }
 \email{arponpodder.ap@gmail.com}

 \author{Mahiratul Jannat}
 \affiliation{%
   \department{Department of Women and Gender Studies}
   \institution{University of Dhaka}
   \city{Dhaka}
   \state{}
   \country{Bangladesh}
 }
 \email{nmahiratul@gmail.com}

 \author{Robert Soden}
 \affiliation{%
   \department{Computer Science and School of the Environment}
   \institution{University of Toronto}
   \city{Toronto}
   \state{Ontario}
   \country{Canada}
 }
 \email{soden@cs.toronto.edu}

 \author{Bryan Semaan}
 \affiliation{%
   \department{Department of Information Science}
   \institution{University of Colorado Boulder}
   \city{Boulder}
   \state{Colorado}
   \country{United States}
 }
 \email{bryan.semaan@colorado.edu}

 \author{Syed Ishtiaque Ahmed}
 \affiliation{
   \department{Department of Computer Science}
   \institution{University of Toronto}
   \city{Toronto}
   \state{Ontario}
   \country{Canada}
 }
 \email{ishtiaque@cs.toronto.edu}


\newcommand*{\rifat}[1]{\textbf{\sffamily{\textcolor{red}{[#1 -- Rifat]}}}}
\newcommand*{\dipto}[1]{\textbf{\sffamily{\textcolor{violet}{[#1 -- Dipto]}}}}
\newcommand*{\ishtiaque}[1]{\textbf{\sffamily{\textcolor{blue}{[#1 -- Ishtiaque]}}}}

\renewcommand{\shortauthors}{Mohammad Rashidujjaman Rifat* and Dipto Das* et al.}

\begin{abstract}

Despite significant research on online harm, polarization, public deliberation, and justice, CSCW still lacks a comprehensive understanding of the experiences of religious minorities, particularly in relation to fear, as prominently evident in our study. Gaining faith-sensitive insights into the expression, participation, and inter-religious interactions on social media can contribute to CSCW's literature on online safety and interfaith communication. In pursuit of this goal, we conducted a six-month-long, interview-based study with the Hindu, Buddhist, and Indigenous communities in Bangladesh. Our study draws on an extensive body of research encompassing the spiral of silence, the cultural politics of fear, and communication accommodation to examine how social media use by religious minorities is influenced by fear, which is associated with social conformity, misinformation, stigma, stereotypes, and South Asian postcolonial memory. Moreover, we engage with scholarly perspectives from religious studies, justice, and South Asian violence and offer important critical insights and design lessons for the CSCW literature on public deliberation, justice, and interfaith communication.

\end{abstract}


\maketitle

\section{Introduction}
\label{sec:introduction}
Despite the initial promise of social media platforms to foster the coexistence of diverse communities, ensure equality of voices in public deliberation, and provide a safe space for participation, these platforms have become unsafe spaces for minority communities -- the religious or ethnic groups that constitute a smaller proportion of the overall population within a particular society or region~\cite{jackson2005minority}. Over the past decade, as social media platforms gained popularity, the issue of inequality and online harms based on cultural, faith, and identity-related factors, including race~\cite{ogbonnaya2020critical, smith2020s}, caste~\cite{vaghela2021birds}, gender~\cite{bardzell2011towards, scheuerman2020we}, and religion~\cite{mizan2019silencing, rifat2020religion}, has received significant attention in academia and public discourse. Academia and the media have reported on how social media platforms have introduced and propagated new forms of online violence targeted at religious minority communities in the Global South~\cite{Human_Rights_Watch_2021, Hasan_Macdonald_Ooi_2022}. For example, over the past few years, a common pattern has emerged in which rumors defaming the holy books of Islam are spread from profiles pretending to be Hindu individuals, resulting in mob attacks on Hindu religious establishments and homes in Bangladesh~\cite{Hasan_Macdonald_Ooi_2022, Ethirajan_2021, VICE_2021}. During the COVID-19 pandemic, fake videos claiming that Muslims are deliberately spreading the virus went ``viral'' on social media platforms in India~\cite{bhatia2022discursive, Human_Rights_Watch_2021, Regan_Sur_Sud_2020, The_Washington_Post_2020}. These false videos led to physical abuses of Indian Muslims and hindered their businesses, as many people boycotted them~\cite{Human_Rights_Watch_2021}. In Bangladesh and other South Asian countries, organized misinformation tends to focus on religious sentiments, unlike in North American regions where misinformation is primarily targeted at politics and electoral systems~\cite{haque2020combating}. Rumors, fabricated images, and targeted memes are frequently posted from social media profiles impersonating members of minority communities, with the aim of inciting social violence South Asian countries~\cite{haque2020combating, kaur2018information, dash2022insights, minar2018violence}. Several incidents of social media-incited mob violence have resulted in the killings of individuals from minority communities~\cite{Hasan_Gooding_Dayant_Johnston_2017, roy2023sociological, al2021content}. In the extreme examples, social media has been found to be complicit in incidents of ethnic cleansing~\cite{Amnesty_International_2023, The_Guardian_2021}. These forms of online harms and violence, magnified through social media, have alarmingly escalated, making the lives of religious minority communities worldwide unsafe due to the instilled feelings of fear, intimidation, and anxiety.





While CSCW and related disciplines have made significant contributions to understanding and mitigating various forms of online harms for racial, linguistic, and other forms of minority communities, many of which are primarily based in the West~\cite{ogbonnaya2020critical, pyle2021lgbtq, erete2021can}, research in this domain is still limited by a lack of understanding of the contextual, social-psychological, and political aspects of social media use experienced by religious minorities in South Asian regions. We argue that studying the social media use and the unsafe experiences of religious minorities in South Asian countries separately from other forms of minority experiences, such as the well-documented CSCW literature concerning racial minorities in the West~\cite{ogbonnaya2020critical, rankin2021black, musgrave2022experiences}, is important for two primary reasons. First, whereas racial minorities' identities are primarily built on individuals' ancestral and genetical background~\cite{helms1994conceptualization, mcintyre1997making}, religious minorities' identities are based on their beliefs and social practices~\cite{fox2000religious}. Factors such as the perceived threat posed to the worldview of majority groups by the actions of minority groups, the use of religion in political legitimacy, and the influence of religion in ethnic conflicts strongly influence the experiences of ethno-religious minorities~\cite{fox2000religious}, which cannot be accurately characterized by discrimination faced by ethnic minorities alone. Second, unlike the race-based social justice concerns that are comparatively more prevalent in most of the West, social justice concerns in South Asia often revolve around faith-based issues~\cite{nandy1988politics, das2001remaking, rifat2020religion, mim2021gospels}. Religion has played an influential role not only in state politics in South Asian countries but also in socio-cultural politics in everyday life, due to its deep intertwining with the history and culture of the region~\cite{robinson2000islam, nandy1988politics}. These unique dynamics may not be fully captured when analyzing minority experiences where race, language, gender, or other factors have been the center of attention. 


Despite these differences, modern human rights frameworks tend to collapse all forms of minority rights under the single umbrella of minority protection rights~\cite{ghanea2012religious}. Moreover, when religious minorities face discrimination, their cases are often handled under the category of ``freedom of belief or religion'' instead of minority rights~\cite{ghanea2012religious}. CSCW literature shows a similar pattern, where religious aspects of minority experiences are often collapsed into other cultural and generalized minority categories, as evidenced by a substantial body of work where religion is considered in the periphery of analyzing online harms (see, for example~\cite{shahid2023decolonizing, mustafa2020patriarchy}). For these reasons above, we argue that the religious aspects of minority experiences are not merely a reflection of cultural, ethnic, and racial experiences but require distinct and separate attention to fully understand the safety and vulnerability of religious minority communities on social media.


To this end, we conducted an interview-based study involving three Bangladeshi religious minority communities to comprehensively understand their experiences of using social media platforms. Informed by our findings, we place ``fear'' at the center of our analysis when examining the experiences of religious minorities in their use of social media in Bangladesh. Our findings reveal three dominant patterns: first, the constant political and social psychological influence of fear within religious minority communities on social media platforms; second, the imposition of fear onto minorities by majority groups and social institutions; and third, the continuous adaptation of social media use by the communities based on their anticipation of future fearful consequences. Our focus on fear is also informed by the long history of the politics of fear concerning religious minority communities in South Asia. The strong religious influence on nationalist politics in the Indian subcontinent has utilized the politics of fear to justify anti-minority violence in the region~\cite{anand2016hindu}. The online harms incited by social media often align with hyper-nationalist majoritarian political ideologies in South Asian countries~\cite{dash2022insights}. For example, in India, Hindu nationalist politics known as Hindutva perceives religious minority groups as a threat to their identity, thus creating a stereotypical image of these minority groups as dangerous~\cite{anand2016hindu, shani2021towards}. Due to these perceived threats, the majority religious groups portray minority religious groups as ``anti-national'' \textit{Others}~\cite{dash2022insights, islam2007imagining, oommen1986insiders}. Similar strong intertwining of religions and politics continue to fuel the politics of fear in other parts of South Asia as well~\cite{mohsin1984religion, riaz2016bangladesh, fair2012faith}. This politics of fear socially constructs hate against religious minority communities in South Asia, and justifies social surveillance, control, and policing against them~\cite{anand2016hindu}. As a result of the amplification of existing societal marginalization of minority communities, as well as the use of social media to incite violence against these communities, social media platforms evoke fear, anxiety, and other negative emotions among minority communities in South Asia~\cite{saha2021short}. Against this backdrop, there is a lack of insights within the field of CSCW and related disciplines regarding the negative experiences of religious minority communities with social media platforms and the support they require. While some aspects of CSCW work in privacy, sensitive disclosure, cyber security, and online harassment have alluded to fear, none are adequate to understand the fearful experiences of religious minorities as informed by our study. Such insights are crucial for understanding and designing strategies to mitigate harms for religious minority communities in the Global South. With this goal in mind, we make the following contributions to the CSCW community:


\begin{itemize}
    \item We contribute to the CSCW literature on online safety by providing in-depth insights into the political and socio-psychological aspects of fear among religious minorities in Bangladesh who use social media. Drawing upon the theories of the spiral of silence, politics of emotion, and communication accommodation, we examine the sources and consequences of fear within these communities. Additionally, we discuss the social conditions that either empower or silence these minority communities through the use of the politics of fear. Finally, we explore how these communities address fear by reorienting their social and online connections and adjusting their participation strategies in public spheres.
    \item We contribute to the existing CSCW literature on polarization, restorative justice, and interfaith coexistence on social media by drawing on insights from religious studies, social justice, and interfaith communication. In our discussion, we explore the aftercare and cultural appropriation of restorative justice as means to enhance online safety for religious minority communities in South Asia. By analyzing our findings in relation to the existing literature on harms and violence in South Asia, we uncover the potential of incorporating considerations of history, memory, and postcolonial trauma in the design of solutions for addressing fear in the minority communities.

\end{itemize}

We believe our study will advance CSCW's broader vision of social justice and online safety.

\section{Literature Review}
We draw from three lines of scholarship to inform and frame our work: firstly, we explore the social and political psychological aspects of fear relating to group coexistence, ethnic conflicts, and peacebuilding. We then extend our discussion to two related lines of CSCW literature and demonstrate how the literature might benefit from addressing fear in the use of social media by religious minority communities. These two areas include the fear-based marginalization of religious minorities in public spheres, particularly social media, and addressing fear in supporting religious minority communities, with a specific focus on restorative justice.


\subsection{Fear and Group Coexistence: Social and Political Psychological Aspects of Fear}
\label{fearandgroup}

The literature of fear defines it as an apprehension of harm to the well-being of a group~\cite{robin2004fear, furedi2006culture, jarymowicz2006dominance}. These apprehensions can vary from panic over crimes to everyday intimidation~\cite{robin2004fear}. Theologian and philosopher Søren Kierkegaard explains the role of fear in creating and dealing with political crises and uncertainties in everyday life ~\cite{kierkegaard2013fear}. Consequently, fear often arises in response to civic crises, acting as an adaptive mechanism for coping with adversity~\cite{jarymowicz2006dominance}. Conversely, fear is often detrimental, limiting freedom and exacerbating group conflicts.

Fear undermines inter-group coexistence and peaceful conflict resolution~\cite{lake1996containing, lake1998international, melander2009geography, thiong2018politics}, leading to mistrust and insecurity~\cite{jarymowicz2006dominance, neocosmos2008politics}. This is particularly true for minority communities, where fear arises from both rational concerns and political myths~\cite{lake1996containing}. These detrimental impacts of fear on group coexistence and peacebuilding are evident in celebrated South Asian scholar Veena Das's analysis of the conflict between Hindu and Sikh communities, where fear stemming from violent memories and traumatic events hindered peaceful negotiation and conflict resolution between the two religious groups~\cite{das1998specificities}. Understanding these group dynamics shaped by fear is crucial for political decision-making and fostering societal peace, affecting compromise and support within groups~\cite{worchel2008between, abu2001reconciliation, gayer2009overcoming, halperin2011emotional}. For example, in ethnic and religious peace negotiations, fear hinders concrete actions and mutual risk-taking~\cite{halperin2011emotional}, exacerbating conflicts by promoting conspiracy theories~\cite{hebel2022vicious}. Overall, recognizing fear's role in minority-majority group coexistence is essential in CSCW, where we provide rich insights on fear-based marginalization on social media and its relevance in advancing restorative justice for religious minorities.

\subsection{Fear-based Marginalization of Minority Voices in Public Spheres}

To examine the impact of fear on social media, we begin by tracing the concept of the public sphere, its foundational principles of freedom of expression and participation, and the challenges it faces in ensuring equal participation due to fear. Subsequently, we delve into the literature of CSCW focused on public deliberation, illustrating how fear can contribute to critical discourses and the design of socio-technical systems aimed at addressing fear within religious minority communities in South Asia.

\subsubsection{Fear's Influence on Public Spheres}
One influential work that has shaped the discourse on free speech, equal public participation, and deliberation is Jürgen Habermas's concept of the public sphere. Habermas describes the public sphere as a space for forming public opinion through rational debates among free, equal citizens~\cite{habermas1996public}. He notes that citizens in this sphere freely express and publish opinions on general interests~\cite{habermas1974public}. Scholars argue this sphere should foster equality and inclusivity, allowing opinion expression regardless of social status~\cite{kruse2018social, shirky2011political, bruns2015habermas}. Social media's features, like free information access and unrestricted participation, align with the Habermasian characteristics of public sphere~\cite{shirky2011political}, making them digital counterparts to the physical public sphere~\cite{ccela2015social}. This led to hopes that social media would enhance equality and inclusivity by amplifying diverse voices.

However, scholars are increasingly becoming skeptical of the social media's capability to enable equal participation~\cite{kruse2018social, polan1990public, kellner2000habermas, dahlberg2005habermasian, susen2011critical}. Studies show that fear, driven by surveillance, media misperceptions, and fear of isolation, silences minority voices on social media~\cite{hampton2014social}. For example, Muslim and Black communities often self-censor online due to law enforcement scrutiny, even in free societies~\cite{aziz2020fear}. This leads to self-monitoring to conform with majority norms~\cite{aziz2020fear}. Negative media coverage further alienates minorities, hindering their social media participation~\cite{hildreth2015fear, fox2015queer}. Studies conducted by Pew Research supports this, showing that individuals with minority identities avoid expressing opinions to evade ostracism~\cite{hampton2014social}. These insights demonstrate fear's role in marginalizing minority voices on social media platforms. We will next explore CSCW research on public deliberation and polarization, highlighting the importance of addressing fear to reduce marginalization of religious minorities in South Asia.

\subsubsection{Public Deliberation and Marginalization of Minority Voices in CSCW}

CSCW literature explores online polarization on social media and strategies to mitigate it. Nelimarkka et al. note that social media often leads to engagement with like-minded people, causing political polarization, and suggest counteracting this by presenting different viewpoints~\cite{nelimarkka2018social}. Kim et al. introduce StarryThoughts, a platform for expressing and exploring diverse opinions, and present their findings about how the authors' demographics on posts enhance empathy and openness to other opinions~\cite{kim2021starrythoughts}. Wang et al. use Moral Foundation Theory to present ideologically diverse news content, helping users understand moral differences expressed in public spheres~\cite{wang2022designing}. Bail et al. show that intergroup communication fosters effective deliberation and mutual understanding in politics~\cite{bail2018exposure}. In summary, the aforementioned and other works~\cite{semaan2014social, terren2021echo, baughan2021someone, das2021jol} within CSCW explore design and policy suggestions, ranging from exposing users to diverse political perspectives to utilizing compromise and mutual understanding as design resources. These studies assume contexts supportive of democratic voices across identities, unrestricted expression, legal safety, and modern rational values.

Contrary to these insights and their underlying assumptions, a line of scholarly studies about South Asian religious minorities shows how the already existing divides between religious majorities and minorities are magnified on social media, resulting in fear and anxiety within the religious minority communities~\cite{bahfen2018individual, bhatia2020examining, bhatia2019times}. In South Asia, religious divides in society drive people to social media for solidarity with like-minded individuals~\cite{bhatia2019times}. However, expressing religious identities online often exacerbates these divides, silencing minority voices~\cite{bhatia2019times}. Intersectional identities, combining multiple social facets, further marginalize minorities~\cite{bhatia2020examining, cho2013toward}. For instance, Muslim women, as minorities in patriarchal societies, face unique challenges in their everyday interactions~\cite{bhatia2020examining}. Even within same religious groups, such as Muslims with non-traditional views, internal conflicts exists~\cite{bahfen2018individual}. These insights highlight the need for CSCW to address religious identity-based marginalization to empower minority voices online and in public spheres.

Combining CSCW research on public deliberation and polarization with studies on the marginalization of religious minorities in South Asia, we identify a key gap: the role of negative emotions and experiences felt by religious minorities on social media platforms, especially in the Global South where socio-political and religious factors are deeply interlinked. Our study aims to address this gap.

\subsection{Design Justice for Minority Communities: The Case of Restorative Justice}
The political and social psychological aspects of fear, as discussed in section \ref{fearandgroup}, together with our findings, indicate the influential role of fear in shaping the pursuit of justice and the sustainability of justice practices for minority communities. Recognizing this pivotal role, our contribution aligns with a separate strand of literature in CSCW that utilizes restorative justice to support victims from minority communities. To contextualize our contribution within this literature, we initially explore the negative experiences encountered by minority communities, followed by an examination of coping mechanisms and available support mechanisms for addressing these negative experiences. Finally, we focus on the CSCW literature concerning restorative justice and discuss the necessity of addressing the fear experienced by religious minority communities in South Asia.

\subsubsection{Negative Experiences of Minority Communities on Social Media}
CSCW has extensively documented the various forms of negative experiences and challenges that minority communities face online. The ``default publicness,''  an inherent design orientation that prioritizes state-validated identities and favor majoritarian views~\cite{cho2018default}, creates disparities between the platform values and the values of the minority communities~\cite{devito2021values}. Because of such disparities, as CSCW literature has examined, members of LGBTQ+ communities face incident of unexpected exposure of their posts to unintended audiences~\cite{cho2018default}, abuse~\cite{scheuerman2018safe}, and stigma~\cite{scheuerman2018safe}. CSCW research has similarly documented emotional and physical discomfort, embarrassment, lack of belongingness, harassment, and demonization, among other harms and unsafe experiences among racial minority groups~\cite{ogbonnaya2020critical, smith2020s, cunningham2023grounds, adamu2023no, salehi2023sustained}. The intersectionality of multiple minority identities further exacerbates the harms experienced by these communities. For example, CSCW has documented how he dual minority identities of transgender women~\cite{scheuerman2018safe}, trans individuals with eating disorders~\cite{feuston2022you}, LGBTQ young individuals~\cite{10.1145/3173574.3173683}, and gender or sexual minorities experienced pregnancy loss~\cite{andalibi2022lgbtq}, among other intersectional identities experienced intensified abuse and marginalization. Collectively, the studies on minority communities confirm that individuals with minority identities along axes of race, gender, sexuality, and more, encounter a range of compounding harmful experiences, including harassment, marginalization, bias, and violence. However, the CSCW literature currently lacks comprehensive insights (with a few exceptions) into the experiences of religious minority communities in Global South.

\subsubsection{Coping Up with Negative Experiences and Seeking Justice}

To cope with the negative experiences and navigate other online safety-related issues, minority communities often find social media platform policies misaligned and sometimes hostile towards them. For example, minority communities are often forced to conform to universal platform policies, leading to further marginalization~\cite{shahid2023decolonizing, feuston2020conformity}. Due to the social media platforms' universal content moderation policies, minority communities lose personal content more frequently than majority groups, even when the content follows community guidelines~\cite{feuston2020conformity, haimson2021disproportionate}. The platforms lack resources to filter out content that is made for ridiculing and bullying minority communities~\cite{medina2020dancing}, and individuals in the community bear an extra burden in seeking support in these situations~\cite{feuston2020conformity}. Furthermore, the social media platforms lack policies to be inclusive of minority communities by increasing their participation~\cite{saha2019language}. Moreover, moderation policies and support for minority communities are weaker in the Global South, as moderation policies often overlook the context and culture of the population in the Global South~\cite{shahid2023decolonizing}. These findings suggest that social media policies are barely adapted to non-dominant identities, norms, and ways of being. To address these limitations and better support minority communities, CSCW is increasingly drawing on various justice frameworks to integrate them into CSCW design~\cite{xiao2023addressing, erete2021applying}. In this paper, we specifically draw on the framework of restorative justice~\cite{xiao2023addressing} due to its capability to support religious minority communities in addressing fear.

\subsubsection{Restorative Justice for Minority Communities}

A nascent body of literature is focusing on restorative justice to mitigate harm for minority communities~\cite{xiao2023addressing, xiao2022sensemaking, chatlani2023teen, rabaan2023survivor, sultana2018design}. Social media companies often address online harm following a justice model primarily centered on ``isolation and punishment''\cite{myers2018censored, vaccaro2020end}. While social media platforms offer users the ability to customize, report, delete, and block content as well as contest moderation decisions, these features effectively place the burden on users, especially those with minority identities, while the platforms hold the power in their hands. In contrast, proponents of restorative justice within CSCW argue for turning the attention from platforms to the vulnerable and affected communities and supporting them\cite{salehi2023sustained, xiao2023addressing}. Restorative justice prioritizes the victims of online harms by restoring communal bonds and relationships~\cite{xiao2023addressing}. Alongside modern values of freedom, equality, accountability, and deliberation, concepts like healing, apology, social harmony, and other human virtues are receiving attention in addressing online harms through restorative justice~\cite{xiao2023addressing}. While these directions offer hope for supporting the victims, as we will argue below, the literature needs to contextualize restorative justice to account for the social-psychological and political aspects of fear in South Asia.

Religious sensitivities, which are more prevalent in South Asian everyday life and national politics, reinforce discrimination and fear against religious minority communities~\cite{bhatia2021religious, bhatia2022hindu, mohan2015locating}. Far-right religious nationalists in South Asia employ social media to perpetuate social fear, discrimination, and stereotypes against minority castes and religions~\cite{vaghela2021birds, bhatia2021religious, mohan2015locating}. Even people who are friendly in society, exhibit hostile and discriminatory behavior online through trolling, cyber-bullying, and other means towards religious minority communities~\cite{bhatia2021religious}. The politics of fear is often supported and normalized by political parties that benefit from such narratives~\cite{bhatia2022hindu}. Overall, religious minority groups are depicted as both perpetrators and victims of online violence~\cite{kirmani2008competing, bhatia2022revolution}. In this context, we emphasize that fear adds an extra layer of challenge in supporting victims through restorative justice. To address this gap, we draw from our findings to provide design and policy implications to address fear and better support religious minority communities in the Global South.

\section{Research Site: Marginalization of Religious Minorities in Bangladesh}\label{research_site}
For a broader readership of this paper, in this section, we will briefly describe how the religious minority communities in Bangladesh often face marginalization. During the late period of British colonization, as a consequence of prolonged colonial ``divide-and-rule" policies, the communal sentiment between Hindus and Muslims in the region was adversarial. Several large-scale communal riots, like the Noakhali riots and The Great Calcutta Killings, took place in 1946, causing the deaths of thousands of Hindus and Muslims~\cite{batabyal2005communalism, sarkar2017calcutta}. These led to the partition of Bengal based on religion, followed by Muslim-majority East Bengal (present-day Bangladesh) and Hindu-majority West Bengal's annexation to Pakistan and India, respectively, in 1947. During Bangladesh's liberation war in 1971, the Pakistani military particularly targeted and disproportionately tortured Hindus in East Bengal~\cite{anam2013pakistan}.

According to a report published by Minority Rights Group International, oppression of Hindus in Bangladesh has been a constant feature in its history even after its independence from Pakistan, and ``since the beginning of the new millennium, the Hindu population has suffered significantly at the hands of Islamic extremists resulting in their further exodus into West Bengal in India"~\cite{minority2018hindus}. Between 2013 and 2021, Bangladesh saw approximately 4000 attacks on its Hindu minority communities, oftentimes which were incited by rumors of religious offense against the majority community on social media~\cite{ganguly2021bangladesh}. One of the most horrifying instances of such attacks on Bangladeshi Hindu communities in recent times was the week-long series of communal vandalization, killings, torching, and ransacking during Durga Puja (the holiest and largest Bengali Hindu festival) in October 2021. The violence was sparked by online rumors that the Quran\footnote{The holy book in the Islamic religion.} had been desecrated when it was allegedly placed at the feet of the idol of a Hindu deity. A man namely, Iqbal Hossain supposedly from the majority Muslim community\footnote{Bengali names are often religiously coded~\cite{das2023toward}. Irrespective of one's belief, their name often discloses the community they were born in.}, confessed to having placed the Quran at a local Hindu temple~\cite{Report_2021}, which was also seen in a nearby CCTV footage. As the rumor started, without identifying the perpetrator, attacks were launched on Hindu temples, houses, and businesses not just in that locality but throughout the country~\cite{hasan2021minorities}. Similarly, rumors of derogatory and blasphemous comments about Islam or its holy Prophet (PBUH) being posted by individuals from social media accounts with non-Muslim names are exploited by Islamist radicals to attack minority communities in the country~\cite{roy2023sociological}. Communal attacks on Hindus at Nasirnagar in 2016 and on Buddhists at Ramu in 2012 were vivid examples of incidents with such patterns~\cite{amnesty2021bangladesh}. Moreover, the alleged and often falsely accused individuals whose names were used to spread those rumors are subjected to legal and bureaucratic red tape~\cite{jibon2016panicked}.

During times of political turmoil surrounding the national elections, targeted attacks against minority groups like Hindus and Christians historically exacerbate in Bangladesh~\cite{minority2018christians, ittefaq2014attacks, rozario}. In addition to minority organized religions, atheists face marginalization and persecution in the country. Around 2015, several prominent individuals were killed for their critical views on religion, as several others were identified on a ``hit list"~\cite{aljazeeraList2015Endangered, theguardianAmerican2015Atheist}. These attacks carried out by extremist groups on minority religious communities and atheists have led to a chilling effect on the sense of freedom and created a climate of fear among individuals of non-dominant religions and faiths~\cite{shackle2018atheist}.
\section{Methods}
To gain a deeper understanding of diverse social and political-psychological aspects of social media use by minority communities, we conducted an interview-based study. In this section, we discuss participant recruitment for semi-structured interviews, the data analysis methods, and our positionality in relation to this study.

\subsection{Data Collection}
We conducted 20 semi-structured interviews (October 2022 - March 2023) with participants in two Bangladesh cities, Dhaka and Khulna. To recruit participants, we initially used purposive sampling~\cite{suri2011purposeful} followed by snowball sampling~\cite{goodman1961snowball} until we reached theoretical saturation. The primary criteria for recruiting participants were that individuals must identify with one of the following communities: Bangladeshi Hindu, Buddhist, Christian, Indigenous, or atheist. Additionally, they needed to have been active on at least one social media platform for a minimum of one year. We only considered individuals of 18 years or over. We circulated our call for participants on online social networks (e.g., Facebook). We also benefited from word-of-mouth advertisements about our study. After interviewing the first few participants, we started running a snowball sampling~\cite{goodman1961snowball} to recruit more participants until theoretical saturation was met. We summarize the demographic information of our participants (n=20) in Table~\ref{tab:participants}.

Due to the sensitivity of the topic of this study, establishing a sense of trust between the interviewer and interviewees was vital. Therefore, the author, who knew someone personally or had the closest social affinity with them, interviewed that particular participant. We interviewed the participants in person or through audio calls using their preferred mediums (e.g., Zoom, phone calls). Before starting the interviews, we briefly explained key points about participation, data management, and its intended use for scholarly writing. Then, we shared the informed consent document with them and asked permission from them to record the audio of the interviews. Once they read and verbally agreed with the terms in the document, we started the interviews. Participation in the interviews was voluntary, and no compensation was offered. The decision of voluntary participation was based on our participants' recommendations during the initial phase of our study. Before sending a formal invitation, we informally discussed this project with five potential participants and assessed their willingness to participate. They were appreciative of the objective of our research as it gave them a space to share their experiences and expressed a fear that any incentives might make their participation and comments questionable by the religious minority community. They suggested voluntary participation to ensure the perceived authenticity of their comments with us.

\begin{table}[!ht]
    \centering
    \caption{Demographic information of the participants.}
    \label{tab:participants}
    \begin{tabular}{p{1.5cm}p{1cm}p{1.3cm}p{2.5cm}p{2cm}p{3.5cm}}
    \toprule
    \textbf{Identifier} & \textbf{Age} & \textbf{Gender} & \textbf{Education} & \textbf{Occupation} & \textbf{Religious background} \\
    \midrule
        P1 & 25-30 & Female & Bachelor's & Unemployed & Christianity\\
        P2 & 25-30 & Female & Bachelor's & Unemployed & Buddhism and\newline Indigenous\\
        P3 & 20-25 & Female & Higher secondary & Student & Buddhism\\
        P4 & 20-25 & Male & Higher secondary & Student & Hinduism\\
        P5 & 25-30 & Male & Bachelor's & Research assistant & Hinduism\\
        P6 & 25-30 & Male & Master's & Unemployed & Hinduism\\
        P7 & 20-25 & Male & Higher secondary & Student & Hinduism\\
        P8 & 25-30 & Male & Higher secondary & Student & Hinduism\\
        P9 & 20-25 & Female & Higher secondary & Student & Hinduism\\
        P10 & 20-25 & Male & Higher secondary & Student & Hinduism\\
        P11 & 25-30 & Female & Bachelor's & Unemployed & Hinduism\\
        P12 & 25-30 & Male & Bachelor's & Unemployed & Hinduism\\
        P13 & 20-25 & Female & Higher secondary & Student & Hinduism\\
        P14 & 25-30 & Non-heteronormative & Bachelor's & Job holder & Hinduism\\
        P15 & 20-25 & Male & Higher secondary & Student & Buddhism\\
        P16 & 25-30 & Male & Bachelor's & Unemployed & Hinduism\\
        P17 & 20-25 & Female & Master's & Job holder & Hinduism\\
        P18 & 30-35 & Male & Master's & Job holder & Buddhism\\
        P19 & 20-25 & Male & Master's & Student & Hinduism\\
        P20 & 25-30 & Male & Bachelor's & Unemployed & Hinduism\\
    \bottomrule
    \end{tabular}
\end{table}

The interviews started with demographic questions and inquiries about general social media use. We then asked them to discuss their recent online experiences, particularly those involving conflict or discomfort, and the connection of these experiences to their religious identities. The interview later focused on their perspectives regarding the real-life impact of their online interactions. The participants also provided insights on the roles of institutions, religious leaders, misinformation, and intersectional identities in shaping their online interactions, as well as how technology affects their experiences both online and offline.


During the interviews, we took detailed notes about their non-verbal cues. We did not gather any personally identifiable information. All interviews were conducted in Bengali; each session typically lasted an hour. We transcribed and translated the interviews into English and deleted the audio recordings. To further de-identify the data, we removed all identifiers from the transcripts so that those cannot be linked to any participants.

\subsection{Data Analysis}
We conducted inductive coding followed by theoretically guided interpretation for our data analysis. Such combinatory approaches are commonly used in HCI and CSCW research~\cite{das2022collaborative, semaan2016transition, kumar2018uber, rifat2019breaking}. We used MAXQDA, a qualitative data analysis tool, to inductively analyze the data. Four authors collaboratively engaged in the iterative inductive coding of the transcripts at weekly online meetings. In this phase, we identified abstract representations of objects, events, and interactions which repeatedly appeared in the transcripts. We also juxtaposed the notes we had taken during the interviews with the transcripts to complement our data analysis. Some examples of open codes in our analysis were \textit{``monitoring social media users' reaction and comment to recent events of communal tension''}, \textit{``observation of participants' interactions on social media by religious majority users''}, \textit{``paying attention to social media influencers' role during events of communal tension''}, and `\textit{``remembering both online and offline interaction with individuals from the religious majority community''}. Since interview data are contextual, following McDonald and colleagues' recommendations~\cite{mcdonald2019reliability}, we did not report an inter-coder reliability score. Upon generating the open codes, we identified their close relationship and frequent mentions of concepts like fear, surveillance and the role of social media, and communication strategies on social media platforms.

Upon completing our inductive analysis, we found that the emergent themes strongly connected to the political and psychological aspects of fear in social media interactions. Following this analysis, we identified relevant literature that provided lenses for interpreting and presenting our findings. This methodological approach -- beginning with inductive analysis and then engaging with relevant theoretical frameworks -- offered several benefits. The inductive approach allowed us the freedom and openness to capture rich insights into our participants' experiences with social media. These frameworks then enabled us to link our findings to broader discourses, allowing us to meaningfully position our findings within the related literature. Such ``thinking with theory'' approaches~\cite{jackson2011thinking} are frequently utilized in various interpretive methodologies.


The choices of the theoretical frameworks were primarily guided by the themes that emerged, combined with our familiarity with the socio-political contexts shaping the lived experiences of religious minority communities in South Asian regions. In our exploration of appropriate frameworks, a body of literature on violence in South Asia provided useful insights. For example, anthropologist Veena Das's work on subjectivity and violence in India offers a rich commentary on how violence against religious minorities is intertwined with everyday social interactions~\cite{das1985violence, das2000violence}, which affects minority communities physically and psychologically through embodied experiences of fear and intimidation. These findings inspired our choice of Sara Ahmed's theory about the politics of fear to further explain how fear is manifested through individuals and social institutions on social media platforms. To gain even richer insights, we also considered the diverse interpretations of our themes inspired by Jackson and Mazzei's suggestions for qualitative data analysis~\cite{jackson2011thinking}. Followed from these leads and inspirations, the exploration of theoretical frameworks led us to Elisabeth Noelle-Neumann's ``spiral of silence''~\cite{noelle1974spiral, noelle1993spiral}, Sara Ahmed's concept of ``cultural politics of emotion''~\cite{ahmed2014cultural}, and Howard Giles' ``communication accommodation theory''~\cite{giles2016communication}. Neumann's theory explaining how and when individuals tend to remain silent or conform to majority opinions when they perceive their own views to be in the minority due to a fear of isolation or social backlash~\cite{noelle1974spiral, noelle1993spiral} was comprehensive to understanding Bangladeshi religious minority communities' experience in social media and how their voices are silenced. Ahmed's work helped us interpret our participants' views and situate those within broader South Asian socio-psychological and institutional power relationships, as she explains how emotions are not only personal experiences but are also shaped and influenced by cultural norms, power structures, and social expectations~\cite{ahmed2014cultural}. Finally, Giles' theory~\cite{giles2016communication}, which suggests that individuals adjust their communication patterns to either converge or diverge with others, depending on their motivations to either emphasize similarity or difference in social interactions, was useful in understanding the Bangladeshi religious minorities' interaction strategies on social media and offline. Thus, these three theories feed each other in a way that helps us understand our participants' perspectives in relation to the politics of fear in Bangladeshi religious minority communities and how it is mediated through social media platforms. 

\subsection{Positionality and Reflexivity}
Prior research in HCI has highlighted emerging tensions around exploitation, membership, disclosure, and allyship in studying marginalized communities~\cite{liang2021embracing}. Therefore, it becomes immensely important to account for these tensions while moving forward with this study. In understanding the marginalized communities' experiences, the authors’ race and ethnicity reflexively bring certain affinities into perspective~\cite{schlesinger2017intersectional}. All but two (who are Americans) authors of the paper were born and brought up in Bangladesh. Two of them belong to the Bangladeshi religious minority Hindu community, while three authors were born in the Bangladeshi religious majority Muslim community. Six authors of this paper identify as cis-gender men, and one as a cis-gender woman. The paper's authors have research experience in postsecular, decolonial, and postcolonial computing, the relationship between faith and technology, and various marginalized identities in the context of the Global South, especially in Bangladeshi communities. The authors' lived experience and research background motivated them to study how the Bangladeshi religious minority communities' experiences are mediated by technology and put them in the capacity to prioritize the agency of local marginalized communities in computing research.

\section{Findings}
We draw on the theories from social-psychological and cultural studies of fear to report three patterns of marginalization of the minority communities’ social media use: (a) the fear of isolation and the spiral of silence, (b) the politics of fear, and (c) coping with fear and marginalization. The theories provide us with the analytical lenses to analyze our findings that contribute to ongoing discussions in CSCW on identity, faith, and polarization with broader implications of making online spaces safer for marginalized communities.

\subsection{Social Psychology of Fear: The Spiral of Silencing of Minority Perspectives}
People feel increasing pressure to suppress their views when they perceive themselves as being in the minority, as they fear social isolation and exclusion. Based on this idea, Noelle Neumann theorizes the ``spiral of silence''~\cite{noelle1974spiral} that explains the growth and spread of public opinion. According to her, people closely observe their social environment to decide whether their opinion conforms with normative views. People tend to remain silent about opinions if they are likely to face social disapproval, while more likely to voice popular or dominant opinions. This fear of isolation stems from the belief that minority opinions would be seen as deviant and face negative consequences. Over time, mass and social media lead to a spiral effect further marginalizing minority opinion~\cite{gearhart2015something}.

\subsubsection{Quasi-Statistical Organ: The Frequency Distribution of Public Opinion and The Feeling of Harmony}\label{quasi-statistical-organ}

The willingness of our participants to speak up or remain silent is significantly influenced by their perception of the social climate. People self-censor themselves by continually assessing the prevailing trends and how much their own convictions align with those--the mechanism for which is dubbed as the quasi-statistical organ~\cite{noelle1974spiral}. It helps individuals evaluate situations where expressing their views may disrupt social harmony, thereby instructing them to remain silent.

Our findings demonstrate several indicators through which social media users from minority communities develop their quasi-statistical organs. They observe how others use various reaction emoticons and comments expressing support or opposition. They usually feel comfortable expressing their opinion in a moderated civil discourse, whereas identifying stubborn or aggressive users invokes a sense of fear. Participant P7 explained if he would be vocal or remain silent about his views and opinions based on their alignment with majority or minority views of that argument:

\begin{quote}
    I see the situation around me. When you see that you are outnumbered, you will be just listening, you won't like to stay in those arguments. On top of that, arguments do not end on Facebook. Someone may make me a target even after the argument is over on Facebook. Then why should I go into a conflict? I should be even more careful. That is why I don't engage in an argument. [P7, Male, Hinduism]
\end{quote}

While individuals from minority communities monitor social media posts to gauge the amount of support, they are also afraid of their social media activities being subject to public surveillance by majority religious groups. A few of our participants described their experience or knowledge of such cases where unrest was incited by a tip-off from those who were in opposition to minority views. In other cases, surveillance of minorities' is followed by condescending lectures or mass reporting of minorities' posts that aim to reinforce a dominant perspective. Examples of these included justifying vandalization and violence and alleging the victims of religious minority communities as responsible for such incidents. In our long-embeddedness across both these surveilled minority and surveilling majority communities, our participants' experiences of feeling targeted correspond with local incidents (e.g., teachers being physically humiliated and suspended for allegedly ``liking" posts that hurt the majorities' religious sentiments) covered by national media (e.g., ~\cite{bdnews24-2024-bangladesh}).

Contrary to these stories of suppressing minorities' voices on social media, some participants also shared stories of how they received social support in times of such crises and marginalization. While they were concerned that many mass social media users in the past supported the violence or remained silent, they found hope in the emerging support for religious minorities from renowned people whom they described as ``celebrities'' on social media. They also appreciated strong opposition from the majority Muslim community against radicalism. Participant P4 described the social support he received from his friends and acquaintances from the Muslim community:

\begin{quote}
    I really appreciated that when the Hindu communities were victims of the violence of 2021, several of my Muslim friends and seniors messaged me to say sorry out of nowhere. They were not responsible for the violence or involved in anyways with those, and I was not a direct victim of those incidents. Still, they said sorry on behalf of others. They called me over the phone to check in with me. [P4, Male, Hinduism]
\end{quote}

Beyond concrete incidents of communal violence, our participants described how the social media landscape has affected traditional inter-faith cultural practices. Several participants explained this phenomenon using the example of social media debate during their biggest religious festival, Durga Puja. Despite the longstanding tradition of Muslims and Hindus exchanging foods and Muslims visiting Hindu temples, some Islamic clerics preached against this practice, labeling it as a prohibited (\textit{haram}) act and cautioning their Muslim peers to abstain from such cultural practices. Thus, our participants were doubtful about inviting their Muslim friends during Puja, fearing that their sense of intimacy with their Muslim friends would be hampered. Some participants even reported instances of personal attacks on their Muslim friends who visited them during Puja. Though many Muslims on social media amplified such interpretations by emphasizing that visiting a Hindu temple and accepting halal food have no bearing on their own faith and instead foster tolerance and empathy among religious groups, such comparatively liberal voices constitute a minority on social media and are often rudely dismissed by the majority.

The above findings illustrate how religious minorities choose to remain silent or speak out based on the frequency distribution of public opinion. The stronger voices of the majority invoke fear and intimidation, influencing their decisions regarding participation on social media.

\subsubsection{The Fear of Isolation: Seeking Societal Approval and Maintaining Reputation}
The spiral of silence theory presents a social-psychological concept of the public sphere~\cite[p. 61]{noelle1993spiral}. Within this framework, people exist within a society where they are consistently subjected to the scrutiny and evaluation of what she referred to as the ``anonymous court''~\cite[p. 62]{noelle1993spiral}. As people develop their quasi-statistical organs to assess their social environment, the fear of isolation, disrespect, and unpopularity among the dominant group serve as guiding forces, compelling them to conform to the established social norms and actively seek social signals of reputation.

Participants shared their experiences of discussing religion on social media platforms and how it changed later online interactions. Upon constantly being preached about orthodoxy, communal hierarchy, and superiority, and proposals for religious conversion from the majority faith, they got disappointed, stopped talking with them, and lost many social contacts. Hence, they made conscious changes in how they represented themselves to their online and offline social peers to avoid social isolation. Participant P7 explained how some of his friends' self-presentation strategies:

\begin{quote}
    I have some friends whose families are religious--practicing Hindus who observe the rituals sincerely. But when they come to campus, because of being afraid, they try to prove themselves secular in front of [Muslims] and do not participate in the religious programs on campus. They act secular to avoid being singled out or to keep their connection with them [the majority community] safe and sound. [P7, Male, Hinduism]
\end{quote}

Moreover, given the history and contemporary instances of religious marginalization in various ways, several participants were advised by their families not to present themselves differently for the sake of getting along with religious others better. Their families sometimes object to their presence on social media altogether. Therefore, some maintain secret social media profiles and find support through covert social media activities like passive browsing of various posts and venting out to close friends in direct messages. Participant P17 explained how family guidelines affected her response and strategies to find support on social media,

\begin{quote}
    The way I was brought up, there was always this impression that I could not be identified as someone different. There was this time in Bangladesh when being Hindu was risky. There was a lot of violence at some point. From then, we have learned that we cannot say anything out loud on social media. When I see a post, I get hurt, keep that feeling to myself, and just choose not to react. I just go to my friend's inbox and start a conversation about it. [P17, Female, Hinduism]
\end{quote}

While our participants highlighted the importance of various moral and professional support of their friends and compassionate acquaintances from the majority Muslim community as discussed in~\ref{quasi-statistical-organ}, they emphasized how belonging to a majority community brings certain power to speak up on social media that many users from religious minority communities cannot afford. For example, participant P20 expressed his expectation for visible public solidarity and societal approval, saying:

\begin{quote}
    Take the example of the religious debate about eating pork or beef. If you look around you in real life, you will obviously see people supporting you, whatever you decide. But the same people would be silent on social media. It feels like I don’t know them. [Silence] gives encouragement to those who spread hatred. [P20, Male, Hinduism]
\end{quote}

Our participants also emphasized the detrimental impact of stereotypes propagated through social and mass media. These patterned narratives contribute to the silencing of marginalized communities and the conflation of perceptions. Furthermore, rituals and identity are often misunderstood and stigmatized, reinforcing the othering of religious minority communities that we elaborate on next.

\subsubsection{Stereotypes and Unconsciousness of Perceiving the Social Climate and Reinforcement of Spiral of Silence}
The development and spread of both positive and negative stereotypes -- preconceived notions or generalizations about a particular group of people based on their characteristics, traits, or social identities that are accelerated and intensified through social and mass media -- can further contribute to silencing marginalized communities~\cite{gearhart2015something}. People tend to simplify social complexity and subtlety, constructing a simplified social environment model. In this process, individuals often rely on consuming indirect experiences and conflating them with their own perceptions. While minorities are particularly sensitive to what Neumann called the ``climate of opinion''~\cite[p. 161]{noelle1993spiral} or ``silent majority''~\cite[p. 93]{noelle1993spiral}, stereotypes about certain groups may lead them to assume that the majority opinion aligns with those stereotypes. Consequently, the perceived negative climate of opinion creates a more unstable environment for religious minority communities, influencing their willingness to express their views openly, suppressing minority opinions as divergent.

Our participants mentioned how rituals, dietary practices, and materiality of identity become important points in forming cultural stereotypes, stigmatization, and alienation of religious minority communities. There exist people who are genuinely curious about the lifestyles of religious minorities and indigenous communities -- how their houses or places of worship look. However, participants emphasized the negative experiences of being stigmatized based on stereotypes about them. For example, most of our Hindu and Buddhist participants discussed how people often ridiculed and undermined their rituals as worshipping idols without understanding the idea or philosophy behind such practices. Our participants, on several occasions, debated with people from Muslim communities about their religious dietary choices (e.g., Hindus not eating beef or Buddhists and Christians eating pork), where their peers and, on some occasions, strangers from the majority community often tried to convince them otherwise. They also shared how their appearance was often judged from the majority religious perspectives instead of their own and received pressure to convert to Islam due to their beards and negative comments and messages after expressing their opinion on wearing ornaments. These stereotypes and material notions of identity expression often contribute to the othering of religious minority communities. They also mentioned the influence of \textit{hujurs} in spreading negative stereotypes through religious programs. As participant P8 explained:

\begin{quote}
    People from outside of the Hindu faith often talk about Hindu practices. Some people from the Muslim community often say vulgar things about our female deities, often in waz mahfils, which creates discomfort in me. Let's say there are 100-150 people at a waz who are coming out of that event, getting a negative impression about Hinduism, or are influenced to think of Hindus in a negative way. [P8, Male, Hinduism]
\end{quote}

These stereotypes being embedded within the social perception make religious minority and majority communities see each other as intrinsically different, leading to communal distrust. Because of the communal distrust and stereotypes thusly created, in case of incidents like Quran desecration, Muslim communities are influenced to believe that someone from another religion is behind the act without critically considering the possibility of other kinds of politics and the self-interests of someone from their own community. Such hasty speculation about other religions marginalizes minority communities unfairly and disproportionately. P19 described his perspective:

\begin{quote}
    I think the reason behind the conflict [in 2021] was political. If you think about it, whoever belongs to a minority community in a country, usually tries to save themselves, whether Muslims in India or Hindus in Bangladesh. So, in Bangladesh, no Hindu person would dare to agitate the Muslim community. So, I think any politically motivated fundamentalist or someone like that did it. When they started vandalizing the idols, I felt like an inferior person and a constant fear was in my mind. [P19, Male, Hinduism]
\end{quote}

Our participants expect a more positive role from Muslim religious leaders. Like social media microcelebrities' role in shaping sociopolitical discourse in Bangladesh~\cite{das2022understanding, rifat2022situating}, they believe that religious influencers -- the preachers who are widely popular on social media for their Waz~\cite{rifat2022putting} -- have greater social responsibilities in these cases. Especially given the different possible interpretations of religious texts, according to our participants, some preachers try to asses whichever interpretation attracts mass people. While participants appreciated some preachers who highlighted religious harmony through their interpretations, they also expressed concerns about some preachers propagating triggering interpretations for popularity. Participant P5 described his views:

\begin{quote}
    Some [Muslim religious preachers] talk about how their religion is better than everyone else's. There are also some of them who discourage insulting other religions. They talk about differences but encourage harmony [among all religions]. ... However, they are very few in number, say one in a hundred. Different speakers preach different interpretations. ... When it comes to religion, people are overly sensitive and often highlight the negative sides of others. [P5, Male, Hinduism]
\end{quote}

Using Neumann's theory, we understood how individuals in the religious minority community observe their social environment, often feel pressured to suppress their views due to the fear of social isolation and exclusion, and how social media mediates their experiences. These individual experiences are also entangled with the South Asian context that we discuss later in this paper.

\subsection{Politics of Fear and Social Control}
The spiral of silence theory provides insights into the social psychological patterns through which religious minority communities subconsciously or unconsciously regulate their behavior on social media, leading to marginalization. Here, minority communities' fear is facilitated through a power relationship with South Asia's broader social, cultural, and state institutions. To examine this pattern of the political aspects of fear, we turn to Sara Ahmed's influential work, ``The Cultural Politics of Emotion''~\cite{ahmed2014cultural}. We discuss how, besides socio-psychological fear, religious minorities become victims of external fears from the religious and cultural institutions' political power.

\subsubsection{Economies of Fear}
Ahmed discusses how fear has been historically used as an instrument for disciplining, punishing, protecting, and governing the less powerful by civil societies and the people upper in the power hierarchies~\cite{ahmed2014cultural}. Explaining fear as a boundary-defining emotion between the ones that are threatening and the ones that are threatened~\cite{ahmed2014cultural}, she describes how fear originating from power politics and institutions may marginalize a community by creating artificial insecurity, stigma, and stereotypes.

Our participants discussed how majoritarian violence against minorities operates in a transnational context and creates a politics of power in which fear is utilized as an instrument to establish and perpetuate power dynamics. For example, Hindu minorities' sufferings in Bangladesh can be used by communal politicians in India to justify their discrimination against Indian Muslims. Similarly, the marginalization of Indian Muslims or Rohingya Muslims is often used as an excuse by many Bangladeshi Muslims to dismiss the marginalization of Bangladeshi Hindus or Bangladeshi Buddhists. These offline incidents and power relationships in transnational politics instill fear within the minority communities that shape the media narratives and social climate in online interactions, where minorities become hesitant to speak up. Participant P5 shared what his family experienced as a Bangladeshi Hindu minority in the aftermath of a major communal riot in India:

\begin{quote}
    Once, there was a big riot before we were born [As Bangladeshi interviewers and authors, we inferred that he was referring to the riot incited by the Babri masjid demolition in India in 1992]. My cousin told me that the chacha [Muslim neighbor], who had a good relationship with us till two days before, rushed to our house to beat my uncle. The good relationship they had worsened all of a sudden. [P5, Male, Hinduism]
\end{quote}

Our participants also shared incidents where institutions successfully intervened to diffuse tension and the possibility of violence against minority communities. They described amidst the instability and communal sentiment in many districts across the country during such a crisis, local governments and law enforcement authorities in some localities were proactive in ensuring their security and mitigating fear. Our participants discussed state institutions' sincerity and shortcomings and how they felt and thought during contemporary major incidents. However, beyond these major incidents, our participants shared stories of their everyday marginalization, both online and offline, for example, being called a \textit{``Bharoter dalal''}: Indian agent or a \textit{``Malaun"}: a derogatory term used by the Muslim community to denote someone who as a sinner for being a Hindu in Bangladesh. Such everyday experiences of being subject to uncivilized addressing in social media because of one's identity~\cite{sultana2022toleration}, recent major incidents, and past occurrences altogether instill a sense of constant fear, inferiority, insecurity, and alienation in the religious minority communities' memories. Like the history of late colonial British India~\cite{pal2021partition}, our participants described the women from minority communities as being particularly targeted for harassment. Participant P2 shared her memory, saying,

\begin{quote}
    [A]fter the incident of 29th September [in 2012] ... The things that have been destroyed have been rebuilt very beautifully. Since I am from Ramu, I know well those who faced casualties have been helped enough. But even after reestablishment, the fear never really fades away from the mind. ... When they were coming to set fire and giving some slogan that I can clearly remember. We were all helpless in the courtyard, thinking about where we were living! That night is like the worst nightmare in my life! I don't talk about it usually. [P2, Female, Buddhism, and Indigenous]
\end{quote}

Our participants discussed how social media exacerbates their marginalization. Our participants agreed that fundamentalist voices from the majority communities are often louder and more active than the supportive voices on social media and fuel communal violence for political motivations. They mentioned some stories covered by mainstream media in the past years where fake social media accounts with Hindu or Buddhist names were used to post religiously insensitive content and agitate Muslim communities as examples. The perpetrators of these fake social media posts often remained unidentified and unpenalized. Participants described their disappointment about the lack of judiciary actions against such malicious social media identity frauds and their fear of being visible on social media through malicious fake accounts. One participant said:


\begin{quote}
    I avoid sharing or posting anything. It often happened that someone opened a fake ID, pretended to be a Hindu person, and posted things that led to violence against the Hindu community, and nothing happened after that. The person who opened the fake ID is not identified or jailed. In such a situation, those who want to speak up are painted as someone trying to spread chaos, which is why I try to be silent. [P17, Female, Hinduism]
\end{quote}

Due to the fear instilled by social and often State institutions, resulting in the internalization of their fear of openly expressing their views on social media, Bangladeshi religious minorities facing discrimination (e.g., Hinduphobia and Atheophobia) often seek out confined spaces on these platforms to find social support. However, within those spaces, they often face intersectional marginalization from moderators and other gate-keeping users people based on castes and sects.


\subsubsection{The ``Uncertainty'' of Future and the Impression of Fear in the Present}

Sara Ahmed describes that the ``object'' of fear does not necessarily have to be present ``here and now"~\cite [p. 65]{ahmed2014cultural}. Rather, anticipating an uncertain future makes people presently fearful. With the fear of constantly being surveilled and exploited for inciting violence by fundamentalists for political gains, our participants talked about their practices to remain mindful of their social media activities and their accounts' security and access. In Participant P18's words,

\begin{quote}
    When I used to see posts ridiculing my religion, I commented and made my arguments. But later, I realized it was not good for me in the long run. It may instead grow anger towards me. Recently, you have seen many incidents [referring to the persecution of atheist bloggers]. Now, we are even more careful. I am even careful in using public Wi-fi. I always log out from Facebook and Instagram after using them. [P18, Male, Buddhism]
\end{quote}

Our participants' observation on social media of others being objurgated, humiliated, and threatened with death, rape, etc., condition them to rarely engage in discussions on social media regarding their religion-based experiences. Some participants talked about their strategy of separating platforms among different social circles, for example, using Instagram for more candid identity expression and Facebook as their impression management site. Such fear-driven self-representation strategies scale up from individual to family to neighborhood and often upward. Participant P16 observed people in rural areas to be more fearful than the ones living in the capital. They described the self-suppression and cautionary measures taken by the minority communities and also the support from some individuals from the Muslim community in their village:

\begin{quote}
    People were getting together with safety measures to guard the temple. I liked that some Muslims also joined them. However, I could feel the fear. You know Kirtan\footnote{A popular Hindu practice of praising and glorifying a form of divinity, with its origins rooted in a Sanskrit word that signifies narrating, reciting, telling, and describing~\cite{lal2009theatres}.}? People in the temple did not even use a mic. They were doing the Kirtan in a very low voice. [P13, Female, Hinduism]
\end{quote}

Some participants discussed how the present impression of fear and uncertainty about the future as a member of religious minority communities in Bangladesh motivates them to make certain life decisions. For example, one participant mentioned his hope of gaining political support during communal violence as his motivation for joining student politics. A few participants were concerned about the declining\footnote{While the raw number of the Hindu population has increased in the last few decades (from 12.3 million in 2011 to 13.1 million in 2022), the percentage of their population is declining (from 8.54\% in 2011 to 7.95\% in 2022)~\cite{bbs2022preliminary}. Different stakeholders have often used this information to construct different kinds of narratives.} percentage of Hindu minorities in the country. They identified different reasons for this decrease in percentage, such as many from the minority communities being afraid of increasing religious intolerance and leaving the country.

Our participants underscored the importance of institutional support and state intervention in addressing religious minority communities' fear. Similar to international human rights organizations~\cite{riaz2021how, hrw2018bangladesh}, a few of them felt that some recent digital legislation (e.g., the ICT Act) falls short in ensuring religious minorities' freedom of expression online and disproportionately penalizes them than violators from the majority community. Some of them specifically were concerned about the disparity of representation of different religious communities in the national curriculum and the erasure of Bengali Hindu writers' works from textbooks. Hence, they recommended an approach to curriculum design that promotes similarities and interconnectedness of different religious values to resist increasing polarization and religious intolerance and encourage communal harmony.

\subsection{Coping up with the Culture of Fear}
We draw upon the communication accommodation theory (CAT)~\cite{giles2016communication} to illuminate our participants' coping mechanisms with interpersonal and intergroup communication dynamics in response to the social and political psychological aspects of fear associated with isolation, uncertainty, and negative emotions. According to CAT, people adjust their communication behaviors through convergence and divergence strategies. While for convergence, the motivation for communication accommodation includes the desire for social approval, the reinforcement of one's identity, and the aspiration to influence others to align with one's perspectives, in the case of divergence, individuals intentionally differentiate themselves from others in communication to highlight their identity differences. People often balance the desire to belong and fear of isolation using their in-group and out-group positionalities through various linguistic behaviors and encompasses para-linguistic, discursive, and non-linguistic cues based on micro-social factors such as the historical relations between groups, faith, status, demography, and institutional support~\cite{giles2016communication}.

\subsubsection{Divergence}
In the face of these social pressures to change their religions, where our participants' religious beliefs are undermined as false or wrong by a large number of people of another religion, they often communicate through divergence. Instead of submitting to the majoritarian views, minorities like Participant P1 try to distinguish and reaffirm their religious identities:

\begin{quote}
    Several times, a few people told me that the basis of Christianity is false. They believe in Jesus Christ in a different way than the whole belief system of Christianity. They tried to say theirs was the only true religion and had answers to everything. All other religions are wrong. Once, when I was little, probably at age 15, I was talking to a guy, and he made me say a Surah\footnote{The equivalent of ``chapter'' in the Quran, which is the sacred scripture of Islam.} [mentions the Surah]. ... Then he said you became Muslim. I asked him, ``How have I become Muslim?''. He said that you just said the Surah. Then I said, how can I be Muslim if it does not come from my heart? [P1, Female, Christianity]
\end{quote}

These divergence attempts need deep religious knowledge, for which our participants look into their religious texts, search for references using the internet, ask questions to their family members, or reach out to religious leaders (e.g., fathers at local churches). They also mentioned people's strong attachment to their beliefs and communities. For example, a social acquaintance of participant P7 posted about the history of historic communal riots and how those marginalized Bengali Hindu communities and a Muslim friend blamed the Hindus for those riots. Several participants described fact-checking as a ``fruitless'' effort and such divergence-based communication as a ``vicious cycle'' in such cases because of people's cognitive biases. Nonetheless, they highlighted the importance of distinguishing and reaffirming their religious identities and perspectives, hoping to make small changes in collective perceptions of them within the majority community. However, these efforts from minority communities often face challenges online. Participant P15 describes his experience:

\begin{quote}
    A few years ago, someone posted a meme that ridiculed one of our major festivals, Maghi Purnima\footnote{One of the largest festivals observed by Bangladeshi Buddhists: \url{https://en.banglapedia.org/index.php/Maghi_Purnima}}. I posted a comment there indicating how much problems and ridicule we face from the bigger religious communities in Bangladesh. Following that comment, I received some threats in my inbox; someone told me they would report my account, post my picture to secret religious groups, etc. Following that incident, I stopped posting anything that might trigger anger among other religious communities. My profile was public back then. I made it private after the incident. [P15, Male, Buddhism]
\end{quote}

The participant described their effort to protest someone bullying their community during a religious festival. However, in their effort to reaffirm their religious identity, they were threatened with being silenced using various social media mechanisms. Several other participants shared similar stories of limiting their social media presence, activities (e.g., post, share, comment, react), and privacy settings because of fear induced by such threats and negative comments.

\subsubsection{Convergence}
With adversarial attitudes of many social media users challenging their desire to differentiate themselves by expressing their religious identity safely and freely, many of our participants use convergence--seeking approval or when the cost of conforming is low~\cite{giles2016communication}, to cope with the culture of fear. In addition to discussing their approaches to convergence, they mention that certain individuals or groups of people, often religious preachers, tried to take some action to address the situation, but their efforts were often limited by their locality or reach on social media. These individuals privately reached out to people they knew and either advised them to stay home or urged them to go home from sites of violence. The intention behind these calls was to control the damage caused by the violence as much as possible, but they didn't have significant power or authority to impact the situation substantially. They often try assimilating themselves with others and searching for social support more privately. Participant P17 described her strategy as:

\begin{quote}
    I obviously take a position when I see these kinds of posts. But I don't engage on social media. I learn the post and then take them up to my friends and close circle and discuss with them. Yes, the defaming posts bring anger and sadness. Definitely, if someone talks badly about my religion, it makes me sad. But generally, I don't express it on social media. No one knows who might see my comments. [P17, Female, Hinduism]
\end{quote}

Several participants described experiencing such fear since their middle school days in the forms of pressures for religious conversion, ridicule, and their religious beliefs being undermined. They expressed a kind of fatigue and fear that accumulated over time, for which they eventually decided not to express and surface their religious identity in most cases openly and rarely participate in religion-related discussions online. In contrast to avoidance as a strategy, some participants deemed discussions an important strategy for convergence. However, as we discussed previously, such discussions from religious minorities often face backlash. Hence, our participants emphasized that popular religious preachers can use their social positions to shape a culture of interfaith tolerance and respect. Participant P1 expressed his expectation, saying:

\begin{quote}
    The role of religious leaders in our country is huge. One of the major sources [of influence] is social media. If religious leaders start telling people to be tolerant, it would really help. Most people in our country look up to them, and [religious speakers] are their ideal figures. [Muslim communities] follow [religious speakers]. We rarely saw any major religious leader say to empathize or forbid doing such things. [P1, Female, Christianity]
\end{quote}

Similar to one of the leading authors and a co-author for whom the contemporary communal violence against Hindu minorities in 2021, commonly referred to as \textit{roktakto sharod}: Sanguinary Fall by the Hindus acted as a strong catalyst motivating this study, most of our participants mentioned that incident in their discussions. They applauded some from the majority Muslim community for condemning the violence based on the facts. However, they were also disappointed that instead of recognizing the Hindus as victims of the violence, many people were invested in impression management as a community through lengthy social media posts that often served no purpose as they tended to provide separate statements rather than engaging in meaningful discussions for supporting the victims and holding the perpetrators of that violence responsible. Participant P13 expressed her disappointment, saying,

\begin{quote}
    Rarely do people from the majority community come forward on social media and simply admit that what is happening is an injustice. They take very complex stands. For example, they described their religion's teachings and tried to prove that the person who put the Quran in the temple was not following those teachings and, thus, was not an ``real Muslim.'' They were trying to avoid responsibility as a community by disowning that person. Some were trying to paint that guy as mentally unstable. [P13, Female, Hinduism]
\end{quote}

The majority Muslim community attempted to disown Iqbal Hossain (context described in section~\ref{research_site}) through complex justifications, which can be described as an example of a community attempting to evade collective responsibility for an individual's behavior. However, our participants emphasized that to address the Bangladeshi religious minorities' marginalization, such problematic patterns should not be overlooked. While the apparent cost of ignoring these incidents to achieve convergence might be small for the religious majority communities, the consequences are more expensive for religious minority communities.
\section{Discussion}


\subsection{Restorative Justice to Support Religious Minorities}

We contribute to the CSCW literature on restorative justice by examining the limitations of social media policies that prioritize ``punishment and isolation'' when addressing harms and promoting online safety for religious minority communities. Our study reveals that religious minority individuals hesitate to seek support after negative experiences on social media due to the fear of social isolation and uncertainties about the consequences. Considering these findings, we argue that a restorative justice model for supporting religious minority communities experiencing online harms must consider the social and political psychological aspects of fear. With the goal of creating an environment where they feel assured and comfortable seeking justice, we propose two recommendations that address fear-related contextual sensitivities. Firstly, we advocate for aftercare support for religious minorities, utilizing the indigenous concept of the circle of healing and incorporating spiritual healing practices within religious minority communities. Secondly, we suggest enhancing the trustworthiness of restorative justice by considering cultural shared meaning-making processes.

\subsubsection{Emphasizing Aftercare for Religious Minorities}

Our findings demonstrated that religious minority communities refrain from taking action and seeking support online in the first place due to the social climate of fear. The findings demonstrate how fear hinders the establishment of communal bonds among our participants, impedes open dialogue within the community, erodes trust between religious groups, and fractures solidarity both online and offline. The participants expressed their concerns that by seeking justice, they may  hurt others' religious sentiments, which in turn may jeopardize their future professional careers and expose their families and relatives to social attacks. Additionally, social media platforms exacerbate this situation by exposing their actions to unintended users and attracting unwarranted attention (see also \cite{cho2018default}). This climate of fear not only erodes the sense of belonging and support among religious others but also within their immediate social circles, including their families. However, these feelings of belonging, trust, and safety are crucial prerequisites for restorative justice to take place successfully, as emphasized by previous studies \cite{zehr2015little, marshall1999restorative}. Furthermore, we note that the intertwining of digital and physical spaces through device sharing in South Asia intensifies the challenges faced by religious minority communities, as highlighted in existing literature \cite{ahmed2017digital, ahmed2019everyone}. Considering these factors, we argue that achieving restorative justice for religious minority communities online necessitates more than merely reaching initial resolutions through discrete actions. It requires the establishment of continuous safety and security measures that provide assurance to members of the minority community, enabling them to overcome their social fears and seek support confidently whenever needed. In essence, addressing fear as an aftercare serves as both a precondition and a facilitator for restorative justice within the religious minority communities.

Drawing on the literature of Indigenous restorative justice, which prioritizes continuous support for victims following initial resolution~\cite{o2007restorative, bazemore2013reintegration}, we propose two suggestions for aftercare in religious minority communities. Firstly, social media platforms can incorporate features similar to indigenous practices known as ``healing circles''\cite{stevenson1999circle, cross2019restorative, heilbron2000traditional}. Indigenous communities have long utilized these circles to foster unity, share stories, build a sense of community, establish deep spiritual connections, and cultivate confidence within a safe space. The literature in theology and mental health further demonstrates that faith leaders from various religions have provided mental health services and mediated such circles, utilizing religious sentiments, scriptural vocabulary, and their professional experiences to create profound and meaningful connections~\cite{daneel2019zionism, osafo2016seeking, green2020traditional}. Socio-technical systems can introduce similar features, allowing victims of online hate within religious minority communities to engage in meaningful conversations moderated by a community leader who already has social acceptance and trust among the community members. These circles not only facilitate healing from the trauma inflicted by fear and attacks on religious identities but also serve as a resource for learning and collectively addressing online hate in everyday social media usage. These suggestions align with the broader call of CSCW to engage with deep spiritual human values, sometimes prompted by faith and spirituality~\cite{toyama2022technology}. Secondly, social media platforms can employ CSCW and computational social science methods to create a safe environment where the posts and actions of minority communities are shielded from undesired exposure to social media users, particularly religious ``others'' who may misinterpret them and potentially incite harm. Implementing these measures ensures aftercare in operationalizing restorative justice for religious minority communities on social media platforms.

\subsubsection{Considering Culture and Processes of Shared Meaning-making in Increasing Trust in Restorative Justice}
Trust plays a crucial role in the acceptance and success of restorative justice~\cite{albrecht2010multicultural}. Building trust and fostering respect among religious groups relies on mutual knowledge sharing and shared processes of meaning-making through cultural practices. However, the meaning-making process becomes challenging when there is a lack of cultural knowledge, leading to stereotypes and prejudices. These stereotypes hinder the effectiveness of restorative justice processes, as they challenge the values and ethics of the cultures of religious others. Our findings highlight how the absence of communication, misconceptions about rituals and ideologies, and myths surrounding the materiality of religious practices among minority communities contribute to the perpetuation of stereotypes held by religious others. We also discovered that simple connections and small conversations can dispel misconceptions, myths, and stereotypes about minority communities. These findings, in line with Albrecht's suggestion~\cite{albrecht2010multicultural}, underscore the significance of interfaith connections, knowledge-sharing, and shared meaning-making of cultural practices in fostering trust in restorative justice among religious minority communities and others. CSCW can utilize methods and techniques from the migration literature, specifically those employing AR to promote intercultural knowledge sharing in multicultural immigrant communities~\cite{sabie2023our}. With this recognition, the next section explores interfaith connections and communication.

\subsection{Addressing Fear in Polarization, Interfaith Coexistence, and Peacebuilding on Social Media Platforms}
Our study contributes to the CSCW literature on public deliberation, polarization, and interfaith coexistence on social media, particularly in South Asia and broadly in the Global South. We analyze the impact of fear in polarized conversations, revealing its role as a barrier for religious minority communities and a political instrument for religious majority communities. Additionally, we provide design and policy implications based on our findings, offering valuable insights for researchers and practitioners in the field.



\subsubsection{Prioritizing Human Virtues for Interfaith Coexistence}
Our findings reveal that fear is utilized by religious majority communities as a political tool to steer interfaith discussions in favor of their own perspectives. Our participants illustrated how many social media users from majority religious groups pretend that they are willing to engage in rational conversations about inter-religious issues, but their acts of engagements show rigidity in accepting any kind of alternative perceptions and beliefs, a phenomena known as superficial ``double monologue'' discussed in interfaith literature~\cite{bijlefeld1996christian, neufeldt2011interfaith}. As an example, our findings on Puja show that Muslim preachers and their followers on social media intimidate Hindus from inviting their Muslim friends during religious festivals. Despite some preachers allowing Muslims to visit and partake in halal food during Puja, rigid preachers and their followers even attack fellow Muslims who accept invitations from Hindu friends. These findings reveal fear and intimidation not only in polarized dialogues between religious minority and majority groups, but also within the same faith groups when minority Muslims do not conform to majority Muslim voices in Bangladesh. In these dialogues, the dominant religious groups often overshadow the underlying political and power dynamics associated with the use of fear-inducing techniques that influence interfaith coexistence.~\cite{10.1145/3555618, rifat2022situating, rifat2022putting}. Such observation show that the online conversation turns into the religious majority's attempt to ``win'' conversations through fear and intimidation, rather than engaging in meaningful discussions.

The role of fear in online polarization, as observed, can be attributed to the disconnect between the ``modern" understanding of deliberation and the theological discourse pertaining to interfaith communication. While ``modern'' conflict resolution and coexistence efforts emphasize free speech and online deliberation, theology suggests that sharing one's beliefs, spirituality, and religious perspectives is qualitatively distinct from ``rational'' communication~\cite{bijlefeld1996christian}. Interfaith coexistence prioritizes human virtues, including compassion, mutual respect, empathy, and trust~\cite{bijlefeld1996christian}. Successful coexistence among faith-based groups cultivates these virtues through verbal and non-verbal exchanges, sharing personal stories, and religious insights within a safe space of a trust circle, allowing individuals to deepen their own spirituality while respecting religious ``others''~\cite{panikkar1999intrareligious, lederach2007reflective, ucko2001thinking}. Building upon our findings and insights from theology, we argue that CSCW design needs to address the patterns of fear and foster deeper connections between faith groups. This can be achieved by designing for human virtues (see, for example~\cite{macintyre2013after}), while also promoting rational deliberation.

\subsubsection{Shared Ethical Interests as a Common Resource to Mitigate Fear in Interfaith Coexistence}
The preceding discussion emphasizes that meaningful inter-religious connections necessitate a safe space that nurtures human virtues, such as trust and compassion. This environment helps to overcome fear and facilitates spiritual connections among faith-based groups in order to address ethical conflicts between them. To achieve this, we propose that CSCW design draws inspiration from the concept of ``coreligionists'' in theology ~\cite{neufeldt2011interfaith, mccarthy2007pluralist}. Coreligionists posses strong social legitimacy, acceptance in the local culture, and aim to establish common ethical languages to help faith-based groups to communicate with religious others ~\cite{neufeldt2011interfaith, mccarthy2007pluralist}. CSCW can utilize computational social science methods to analyze conversations among faith-based groups and explore ethical confluences and conflicts. Our participants mentioned faith leaders both in majority and minority religions that often work as local mediators. These mediators exhibit the similar role as coreligionists and could be utilized by CSCW design to include them to foster meaningful interfaith communication. This could involve developing a platform similar to Trustnet~\cite{jahanbakhsh2022leveraging}, but with careful moderation by coreligionists. Future research could examine how faith leaders from various religious groups can serve as coreligionists, leveraging their existing social capital among their respective followers to help mitigate fear among minority religious communities in South Asia.

\subsection{The Harm Narrative in South Asia: Unspoken Harms, Postcolonial History, Religious Sensitivity, and Justification of Harms}

We contribute to the CSCW literature on negative experiences of minority communities on social media platforms by highlighting a pervasive culture of unspoken harm, where fear leads religious minority communities to silently accept their negative experiences. Our findings reveal two recurring observations: (a) participants hesitated to express concerns due to fear of consequences, and (b) many participants expressed resignation, believing that everyone faces these harms or that speaking up wouldn't make a difference. Instances of undisclosed experiences due to fear were evident through participants canceling interviews, requesting non-sharing of interview recordings, and exhibiting hesitation and avoidance of certain topics. Moreover, participants perceived certain negative experiences as mundane, expressing indifference because they believed others shared these experiences or speaking up would be ineffective. This culture hinders scrutiny and the addressing of fearful experiences.

These findings of unspoken negative experiences and fear resonate with the enduring influence of postcolonial memories and strong religious influences in South Asia, which shape the interpretations, recognition, and justification of harm and violence~\cite{das1985violence, das2006life, das2001remaking, mahmood2010fighting}. Celebrated South Asian scholars, Das and Nandy, illustrate how South Asian revolutionaries, in response to colonial oppression, turned to violence as a means to attain freedom, drawing inspiration from religious traditions that justified violence for a higher purpose~\cite{das1985violence}. This gave rise to a ``structure of ideas'' that legitimized acts of violence and harm in South Asia as commonplace~\cite{das1985violence}. Over time, the language used to describe harm and violence transformed, as bureaucratic language limited the label of violence to incidents that disrupted state law and order~\cite{das1985violence}. The authors assert that bureaucratic language itself is rooted in a history shaped by colonial restructuring and hierarchical categorization of races, cultures, and civilizations.

Our aim in recounting these historical narratives and the legitimization of violence in South Asia is twofold: first, to highlight that the expression, articulation, and justification of harms and violence are influenced by the political climate of fear, together with colonial memories and histories, as well as some forms of religious interpretations and norms; and second, to draw attention to the oversight of religious sensitivities in the policies implemented by the partnership between the local secular Statecraft and social media platforms, which hinders a comprehensive understanding of the underlying factors contributing to the unspoken nature of harms in our study. We call for action to consider the historical and religious context when addressing South Asian harms, which often remain unspoken and unscrutinized among the vulnerable religious minority communities. In doing so, future CSCW research about religious minorities can draw from research methodologies about studying vulnerable communities in sensitive settings~\cite{pacheco2018doubly, shaw2020ethics} to develop nuanced and context-sensitive approaches to tackle the complex challenges faced by individuals and communities affected by online harms in South Asia.

\subsection{Double Consciousness of Religious Minorities and the Veil in Interfaith Communication}
Sociologist W.E.B. Du Bois, in his influential book ``The Souls of Black Folk"~\cite{du1909souls}, introduced ``double consciousness'' to explain how marginalized African Americans perceive themselves in a predominantly white society~\cite{du1909souls, brown2020sociology}. They develop a sense of twoness, aware of their unique experiences and perspectives, as well as the racial stereotypes and marginalization imposed by the broader society. In our study, participants described altering their self-representation based on how they are perceived by the religious majority, such as acting secular or avoiding religious practices in public, based on how they viewed themselves through the lens of how the religious majority communities see them. In those cases, while the individuals from the religious minority communities were aware of their own beliefs and practices, simultaneously, they were also aware of the stereotypes and prejudices commonly held by many in the religious majority community. This creates a division within their identity as they navigate the expectations of the dominant group. Our participants, like African Americans in Du Bois' framework, adopt similar strategies to preserve social relationships and reputation, fearing the consequences of not conforming to majority religious norms.


While the postcolonial partition of Bengal into the Indian state of West Bengal and Bangladesh (then East Pakistan) based on religion is a strong evidence of such identity crises~\cite{chatterjee1993nation}, finding analogies with Du Bois' concepts of double consciousness bolsters the notion of the Bengali people being fragmented in their imagination of communities across religions. Based on our study and recent events of marginalization of religious minorities in South Asia, we argue that religion serves as the thrust and socio-psychological factor of power in the Indian subcontinent, similar to race in the Western context. Relating to Du Bois' works that emphasizes the need for recognition, understanding, and empathy to bridge the gap between different racial communities, our study highlights how individuals grapple with fear and marginalization while navigating the religious dynamics in their sociocultural settings and urges for better facilitated interfaith communication.

\subsection{Limitations and Future Work}
One primary limitation of our study is the limited age diversity of our participants. While the diversity of religious backgrounds is also limited, it is commensurate with these communities' population distribution in the country~\cite{bbs2022preliminary}. In this study, we have focused on broadly defined religious minority communities in Bangladesh. However, there are various nuances and plural traditions within religious communities in Bangladesh (e.g., Matuas--a Hindu sect primarily comprising of underprivileged castes like the \textit{namasudras}~\cite{lorea2020religion} and Ahmadiyyas--a minority Islamic sect~\cite{uca2023bangladesh}) that also face marginalization in the country in myriad ways due to their intersectional faith-based identity.Our future work will take into account the nuances of identities and plural culture within the same religious tradition, as well as their implications for faith-based marginalization. This paper presents a semi-structured interviews-based study. While interview data are useful for understanding individuals' experiences, we will use qualitative and quantitative content analysis-based approaches in our future work to understand the dynamics of the interaction of religious minority communities around times of national sociopolitical important events. Moreover, our study is based on perspectives shared by a set of participants with somewhat homogeneous demographics. In the case of this paper, similar to the nature of qualitative research~\cite{leung2015validity}, our objective is not to produce generalizability but rather to study a specific process in a defined context.
\section{Conclusion}

This paper presents a study of the political and social-psychological aspects of fear experienced by Bangladeshi religious minority communities on social media platforms. We found that the fearful experiences of these communities are tied to social isolation, pressure to conform to the norms of dominant religious groups, and stereotypical portrayals of non-dominant religious groups. We further examined how power dynamics between religious minority communities and dominant groups, as well as between social and state institutions, perpetuate cycles of fear. Additionally, we investigated coping mechanisms (or the lack thereof) in these communities' fearful engagements with social media. These findings highlight the significant role of fear in the online participation of minority communities and their social and online coexistence with majority religious groups.

Drawing from these findings, we argued that popular justice-oriented approaches, often evoked to support victims of online harms, must consider the social and psychological aspects of fear and create an environment where victims feel safe and are assured of support. We then extend this discussion to address fear in polarized online conversations and explore pathways for interfaith coexistence on social media platforms. One of our recommendations is to engage with the socio-cultural fabric, in our case, South Asian postcolonial society, to recognize unspoken harms and extend support, whether technological or otherwise. Finally, we extend our discussion to the identity and visibility of religious minority communities on social media platforms. We believe this study opens up a range of subsequent research opportunities that can acknowledge and address fear in the design of policies and technologies for online safety in CSCW literature.
\begin{acks}
    We thank our participants for their participation and valuable insights. Special thanks to our anonymous reviewers, whose constructive comments helped to improve this paper. This research was made possible by the generous grants from the Schwartz Reisman Institute Graduate Fellowship (awarded to Mohammad Rashidujjaman Rifat), the Schwartz Reisman Institute Faculty Fellowship (awarded to Syed Ishtiaque Ahmed), the Data Sciences Institute Catalyst Grant, the Natural Sciences and Engineering Research Council of Canada (RGPIN-2018-0), the Canada Foundation for Innovation (37608), and the Ontario Ministry of Research and Innovation (37608).
\end{acks}


\bibliographystyle{ACM-Reference-Format}
\bibliography{main}
\balance




\received{July 2023}
\received[revised]{January 2024}
\received[accepted]{March 2024}

\end{document}